\documentclass[aps,prl,reprint,longbibliography]{revtex4-2}
\usepackage{graphicx}
\usepackage{amsmath}
\usepackage{amssymb}
\usepackage{natbib}
\usepackage{float}
\usepackage{xcolor}
\begin{document}
\title{Multiple moir\'e polarization textures in twisted trilayers}
\author{Haoshen Ye}
\author{Ying Zhou}
\author{Shuai Dong}
\email{sdong@seu.edu.cn}
\affiliation{Key Laboratory of Quantum Materials and Devices of Ministry of Education, School of Physics, Southeast University, Nanjing 211189, China}
\date{\today}

\begin{abstract}
	Twisted van der Waals layers exhibit nontrivial polarization textures from moir\'e-modulated stacking and the inherited sliding ferroelectricity. However, in any twisted bilayer, the topology of polarization texture is essentially fixed due to the inherent twist-stacking landscape for a given material. Here, this constraint is overcome in some twisted trilayers, where the relative moir\'e phase emerges as an additional structural degree of freedom. Using CrI$_3$ and BN as representative sliding ferroelectrics with distinct symmetries, it is clarified how the relative moir\'e phases reshape the trilayer stacking landscapes and tune the net polarization texture by layer sliding. The multiple moir\'e polarization textures can be attributed to the interference of original moir\'e patterns of the two bilayer subsystems.
\end{abstract}
\maketitle

Moir\'e superlattices formed by twisted bilayers provide a versatile platform for realizing exotic physical states absent in the constituent layers, including superconductivity, correlated insulating states, and nontrivial polarization textures \cite{Nat-2018-TBG-Superconduct, Nat-2018-TBG-Correlated, NatCom-2025-BN-Texture}. A key advantage of twisted bilayers is that many of their physical properties can be effectively tuned by the interlayer twist angle, such as the magic-angle condition for superconductivity and correlated insulating states \cite{NatMater-2025-3LGra-MagicAngles, PRL-2024-TMD-MagicAngles}. Likewise, moir\'e polarization textures exhibit a strong twist-angle dependence, with their susceptibility and vorticity enhanced at smaller twist angles \cite{npj2DM-2022-MoS2-Sus, Sci-2024-MoS2-Vortex}. This tunability has made twisted moir\'e materials a central topic in condensed-matter physics and a promising candidate for functional electronic and ferroic devices \cite{NatPhys-2021-Moire-Review, NatRevMater-2024-Moire-Review}.

The polarization textures in twisted bilayers arise from sliding ferroelectricity \cite{ACSNano-2021-Sliding-Review, PRL-2023-Graphene-SlideFE}, where the local dipole is determined by the local stacking mode of van der Waals layers. Twist-induced stacking domains and domain-wall networks give rise to spatially varying polarization textures. For example, meron-antimeron textures are found in twisted bilayers of BN and transition-metal chalcogenides \cite{NatCom-2023-BN-Texture, NanoLett-2024-WSe2-Texture}, and vortex–antivortex textures are formed in twisted CrI$_3$ bilayers \cite{PRL-2025-CrI3-Twisted}. These moir\'e polarization textures carry nontrivial topological charges or winding numbers, leading to robustness against spontaneous structural relaxation and twist-angle variation \cite{PRL-2025-CrI3-Twisted}.

However, such topological robustness limits the switchability of polarization textures in twisted bilayers. In an ideal twisted bilayer, the topology of polarization texture is locked to the underlying texture of stacking modes. Varying the twist angle mainly rescales the moir\'e period, while interlayer sliding translates the moir\'e pattern without changing its topology [see Fig.~S8 in the Supplement Materials (SM) for more details]. Although an external electric field can transiently distort the polarization texture, the system is expected to recover its optimal texture once the field is removed \cite{NatMater-2023-WSe2-Texture, PRL-2025-CrI3-Twisted}. Many efforts have been devoted to pursuing tunability by extrinsically modifying the ideal moir\'e superlattice, such as multiterraced boundary and strain engineering \cite{NanoLett-2026-BN-strain, AdvMater-2025-WSe2-Terraced, ACSNano-2025-WSe2-Srain}. Recently, twisted trilayers and multilayers have emerged as a more complex platform offering multiple interacting interfaces \cite{Nat-2021-Graphene-Twist3L, NatCommun-2024-Graphene-Twist3L}. Interference between multiple moir\'e subsystems and the resulting supermoir\'e phenomena have also been extensively explored in twisted multilayers, primarily in the context of electronic and excitonic properties \cite{JPCM-2025-tTLG-MoMReview, NatCommun-2024-Graphene-Twist3L, NatCommun-2025-WSe2WS2-BichromaticMoire}. In particular, nonvolatile polarization switching was reported in twisted WSe$_2$ trilayers, but the observed switching is closely tied to heterostrain, rather than being an intrinsic property \cite{NatNanotechnol-2024-WSe2-3L}. Thus, whether one can realize intrinsically switchable and nonvolatile moir\'e polarization textures in an ideal twisted superlattice remains a challenging open question.

In this Letter, we theoretically demonstrate that some twisted trilayers can host switchable moir\'e polarization textures beyond the limitation of bilayers. A twisted trilayer can be naturally decomposed into two twisted-bilayer subsystems and a key ingredient emerges, namely the relative moir\'e phase between subsystems. Here, we establish a criterion for identifying energetically favorable relative moir\'e phases and reveal symmetry-governed interference between the polarization textures of adjacent moir\'e subsystems. For generality and diversity, CrI$_3$ and BN are chosen as representative systems with distinct stacking routes to sliding ferroelectricity \cite{PRL-2023-Sliding-Theory, PRB-2025-Sliding-Theory}.

\begin{figure}
	\centering
	\includegraphics[width=0.48\textwidth]{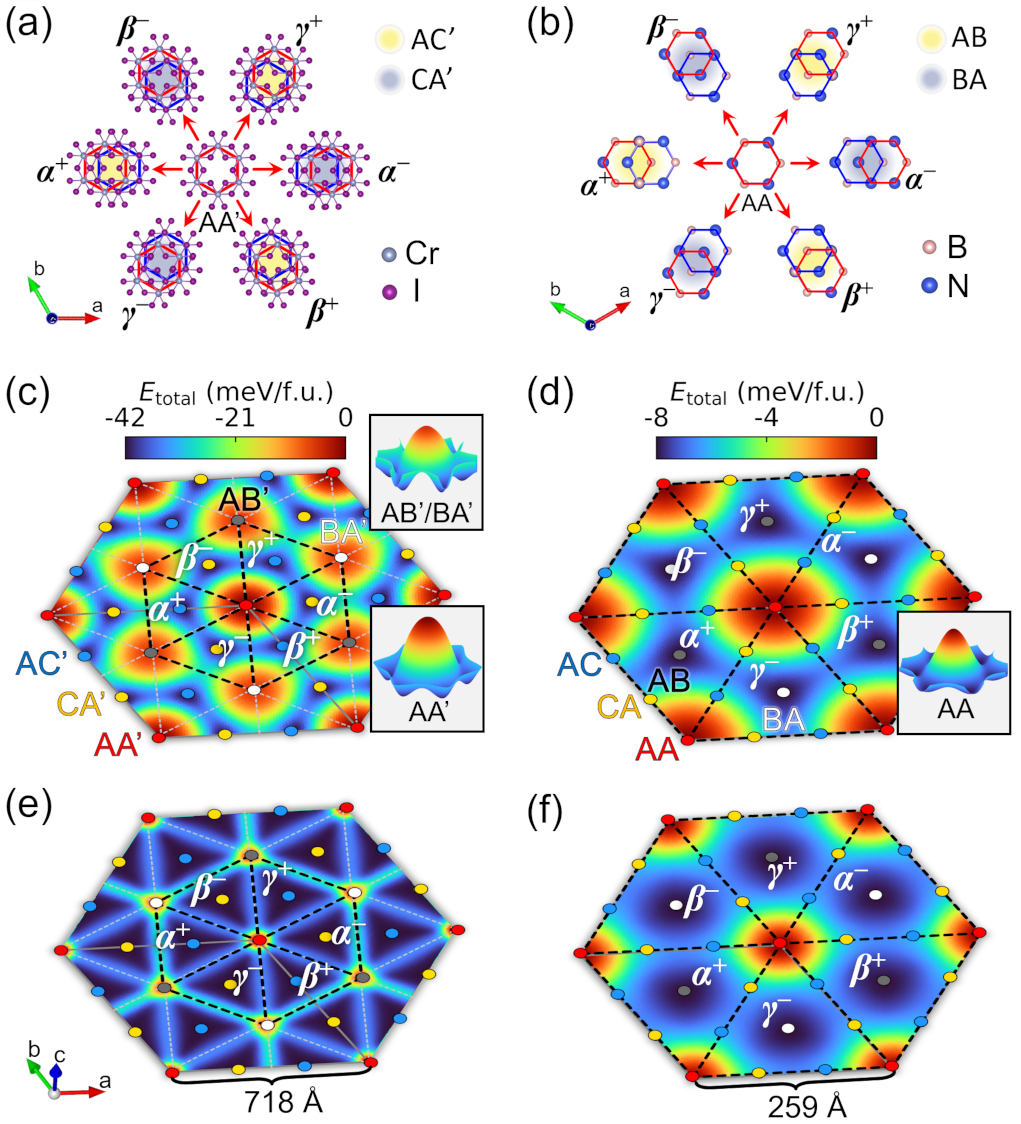}
	\caption{The $\mathbb{Z}_3 \times \mathbb{Z}_2$ structures of (a) antiparallel stacking CrI$_3$ bilayer and (b) parallel stacking BN bilayer, obtained via interlayer sliding. Red and blue frames denote the top and bottom layers, respectively. Red arrows indicate the sliding directions of the top layer, indicated by $\alpha$, $\beta$, or $\gamma$. The superscript denotes the sign of out-of-plane polarization. Local stacking energy ($E_{\rm stack}$) landscapes of $0.56^\circ$ twisted 
		(c) CrI$_3$ bilayer and (d) BN bilayer without structural relaxation, which are obtained by mapping the energy of corresponding sliding bilayers (Figs. S1 and S2 in the SM). Special local stacking modes are labeled together with color dots. Insets: Mexican-hat-like energy landscapes around the high-energy points. 
		Local total energy ($E_{\rm total}$) landscapes of $0.56^\circ$ twisted (e) CrI$_3$ bilayer and (f) BN bilayer after structural relaxation. $E_{\rm total}=E_{\rm stack}+E_{\rm elastic}$, where $E_{\rm elastic}$ is the structural distortion energy of two layers. Details of the structural-relaxation method can be found in the SM \cite{PRB-1996-VASP, PRL-1997-PBE, JCP-2010-vdw, PRB-1998-GGAU, PRB-1993-BP, N-2017-CrI3, S-Song-2018-CrI3, b-strain, PRB-2007-thermal}.}
	\label{Fig1}
\end{figure}

Let us start from sliding bilayers, the core building blocks governing the fundamental physics of twisted layers. Despite the microscopic difference in stacking route [see Fig.~S6 in the SM for more details], both the non-sliding parent structures (i.e. the antiparallel AA' stacking CrI$_3$ bilayer and parallel AA stacking BN bilayer) host the six-fold inversion symmetry $\overline{6}$. Then, any sliding-induced stacking can be decomposed into the $\mathbb{Z}_3 \times \mathbb{Z}_2$ manifold, as shown in Figs.~\ref{Fig1}(a) and (b), where $\mathbb{Z}_3$ sector labels the sliding-direction related variants ($\alpha$, $\beta$, $\gamma$) and $\mathbb{Z}_2$ sector labels the sign ($+$ or $-$) of out-of-plane polarization $P_z$ \cite{NatRevMater-2017-ZmZn-Review, PRL-2012-HMO-Z3Z2}.

Upon twisting, the local stacking varies continuously across the moir\'e superlattice, unfolding the $\mathbb{Z}_3 \times \mathbb{Z}_2$ manifold into a real-space stacking texture [Figs.~\ref{Fig1}(c) and (d)]. Originating from the interfacial interaction, the stacking energy landscape in real space exhibits Mexican-hat-like profiles around all points of high-energy stacking modes. Interestingly, as a general characteristic manifestation of the $\mathbb{Z}_3 \times \mathbb{Z}_2$ manifold, similar Mexican-hat-like energy profile was invoked to explain the origin of topological vortex domain structures observed in hexagonal manganites \cite{NatMater-2014-YMnO3-Z3Z2}. Consequently, the stacking texture forms the vortex-like sequence $\alpha^+ - \beta^- - \gamma^+ - \alpha^- - \beta^+ - \gamma^-$, giving rise to a topologically protected moir\'e structure, similar to the case of hexagonal manganites \cite{NatMater-2010-YMnO3-Z3Z2}.

In realistic moir\'e superlattices, structural relaxation spontaneously occurs to reduce the total energy through the competition between interlayer stacking energy and intralayer elastic energy \cite{NatNanotechnol-2020-TMDs-TwistRelax, NatMater-2021-Graphene-StrainFields}. As shown in Figs.~\ref{Fig1}(e) and (f), structural relaxation enlarges the low-energy stacking regions into domains, while preserving the $\mathbb{Z}_3 \times \mathbb{Z}_2$ vortices and Mexican-hat-like energy landscapes, i.e. the topology of stacking texture is unchanged. Due to the one-to-one mapping between the stacking mode and sliding ferroelectric polarization, a twisted bilayer inherits a unique polarization texture, fixed by its unique stacking-energy landscape. An external electric field can only induce a transient distortion of the polarization texture \cite{PRL-2025-CrI3-Twisted}, rather than switch it to a distinct nonvolatile state.

\begin{figure}
	\centering
	\includegraphics[width=0.48\textwidth]{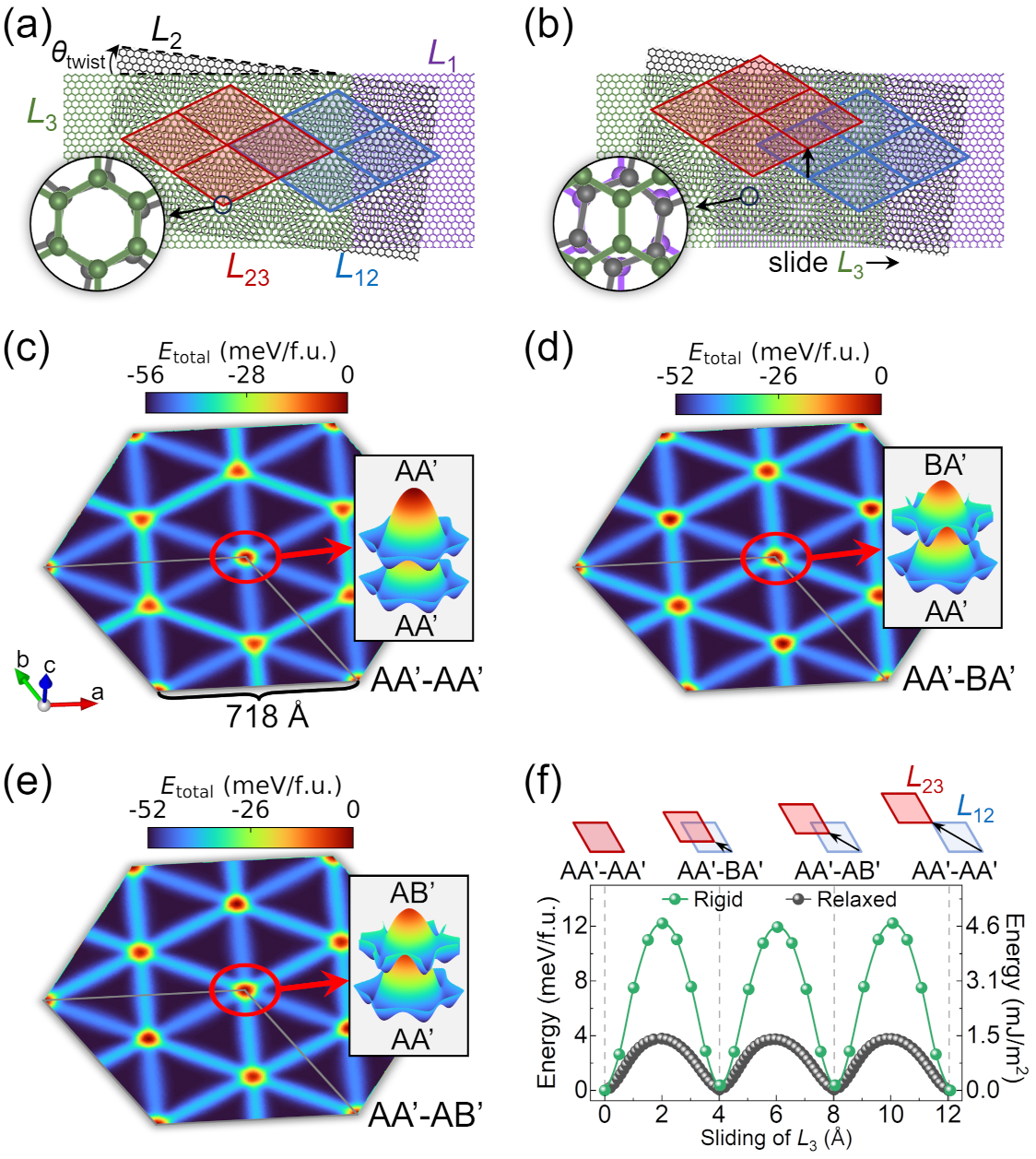}
	\caption{
		(a) Schematic of a twisted trilayer, which can be decomposed into two oppositely-twisted bilayer subsystems, $L_{12}$ and $L_{23}$. The blue and red frames denote the moir\'e periods formed by $L_{12}$ and $L_{23}$ twisted bilayers, respectively.
		(b) Modulation of the relative moir\'e phase by lateral sliding of the top layer $L_3$. Insets show the local atomic structures.
		Local total energy ($E_{\rm total}$) landscape of $0.56^\circ$ twisted CrI$_3$ trilayers for the (c) AA'-AA', (d) AA'-BA', and (e) AA'-AB' states, labeled by the local stacking combination of $L_{12}$-$L_{23}$ at the red-circle regions. Insets show the decomposed energy landscape, which exhibit nested Mexican-hat-like landscapes. 
		(f) Switching barriers along one complete sliding period of $L_3$ obtained from rigid  and relaxed sliding calculations. The schematics above illustrate the variation of the relative moir\'e phase between $L_{12}$ and  $L_{23}$.}
	\label{Fig2}
\end{figure}

In a trilayer, the net polarization texture arises from the superposition of textures from two moir\'e subsystems. For simplicity, here the twisted trilayer is constructed by twisting the middle layer ($L_2$) only, while keeping the top and bottom layers ($L_1$ and $L_3$) aligned, as illustrated in Fig.~\ref{Fig2}(a). Then either $L_1+L_2$ (i.e. $L_{12}$) or $L_2 + L_3$ (i.e. $L_{23}$) alone is equivalent to a twisted bilayer, and the twist angles of $L_{12}$ and $L_{23}$ are opposite. This choice makes the moir\'e periods of $L_{12}$ and $L_{23}$ identical, avoiding more complex superstructures of moir\'e patterns.

Since a moir\'e pattern itself arises from the geometrical interference between two periodic lattices, the coexistence of two moir\'e subsystems in a trilayer naturally introduces interference between their associated textures. When sliding $L_1$ ($L_3$), the moir\'e pattern of $L_{12}$ ($L_{23}$) shifts, while that of $L_{23}$ ($L_{12}$) is unchanged [Fig.~\ref{Fig2}(b)]. Then a relative moir\'e phase emerges between $L_{12}$ and $L_{23}$, which provides an additional structural degree of freedom absent in bilayers. Consequently, the stacking texture is no longer uniquely fixed but tunable by this relative moir\'e phase between two adjacent subsystems. Notably, $L_2$ sliding shifts both moir\'e patterns of  $L_{12}$ and $L_{23}$ without changing their relative moir\'e phase.

Although the relative moir\'e phase can be tuned continuously, the energetic stability of different relative moir\'e phases needs to be analyzed. Since the two adjacent moir\'e subsystems share the same middle layer, their structural relaxations are coupled. An energetically favorable relative moir\'e phase maximizes the spatial overlap of low-energy domains of $L_{12}$ and $L_{23}$, allowing the same middle-layer deformation to reduce additional energy cost. Then, the intuitive expectations are those stackings of bilayer subsystems with nested Mexican-hat-like energy landscape, similar to the Stud-Tube interlocking mechanism of LEGO bricks.

We first analyze the twisted CrI$_3$ trilayer. The twisted CrI$_3$ bilayer contains three Mexican-hat-like energy landscapes centered at the AA', AB', and BA' stacking points. Taking the local AA' stacking of $L_{12}$ as the reference point, there are three states with nested Mexican-hat-like energy landscape in the trilayer system, labeled by their stacking combinations of $L_{12}$-$L_{23}$: AA'-AA', AA'-BA', and AA'-AB', as shown in Figs.~\ref{Fig2}(c-e). These three states are nearly degenerate, with the AA'-BA' and AA'-AB' states lying less than $0.1$ meV/f.u. above the AA'-AA' state for the $0.56^\circ$ twisted cases (Fig.~S13 in the SM). The AA'-AA' state serves as a reference structure with a zero relative moir\'e phase, while in the other two cases the relative moir\'e phases are $\pm2\pi/3$. In all these three states with nested Mexican-hat-like energy landscape, the low-energy stacking domains are predominantly composed of alternating AC'-CA' and CA'-AC' domains surrounding the nested Mexican-hat nodes.

The scenario of nested Mexican-hat-like energy landscape can be verified by calculating the total energy of trilayer as a function of $L_3$ sliding. As shown in Fig.~\ref{Fig2}(f), there are indeed three low-energy states, corresponding to the aforementioned AA'-AA', AA'-BA', and AA'-AB' stacking modes. Switching among these states can be mechanically driven by a lateral force applied to $L_3$, as demonstrated by our force-driven simulations (see SM for more details). For the $0.56^\circ$ twisted case, the force-driven simulations yield a critical force of $0.08$ nN (Fig.~S15 in the SM), and the structural relaxation substantially reduces the switching barrier from $12.2$ to $3.79$ meV/f.u. The critical force is generally reduced at larger twist angles, consistent with a superlubricity-like flattening of the sliding-energy landscape \cite{npjComputMater-2025-HfO2-superlubric, PRB-2004-Graphene-Superlubricity}.

\begin{figure*}
	\centering
	\includegraphics[width=0.65\textwidth]{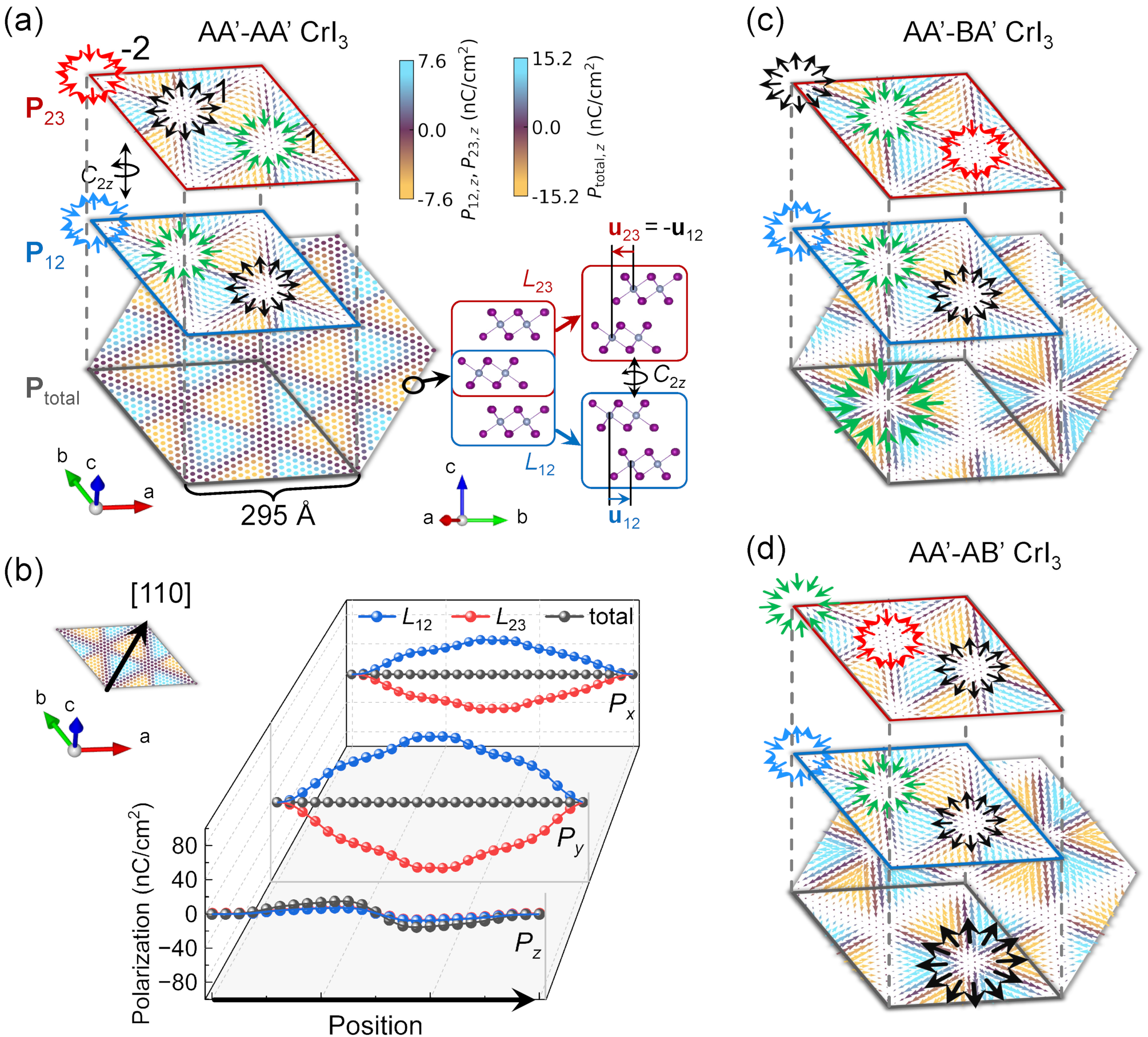}
	\caption{Polarization textures of twisted CrI$_3$ trilayers. Without loss of generality, here a larger twist angle $1.35^\circ$ is used for better visualization. 
		(a) Subsystem-resolved and total polarization textures of AA'-AA' state. The arrow color represents the local out-of-plane polarization $P_{z}$, while the arrow direction and length represent the direction and magnitude of the in-plane polarization. ${\bf P}_{12}$ and ${\bf P}_{23}$ denote the polarization textures of $L_{12}$ and $L_{23}$. ${\bf P}_{\rm total}={\bf P}_{12}+{\bf P}_{23}$ is the net texture of trilayer. The numbers denote the winding numbers of the (anti)vortices. Inset: the  $C_{2z}$ symmetry relationship between $L_{12}$ and $L_{23}$. 
		(b) Polarization components of ${\bf P}_{12}$, ${\bf P}_{23}$, and ${\bf P}_{\rm total}$ along the [110] path indicated in the inset.
		(c-d) Subsystem-resolved and total polarization textures of the AA'-BA' and AA'-AB' states.}
	\label{Fig3}
\end{figure*}

Now we turn to discuss the net polarization texture of CrI$_3$ trilayer, which is naturally determined by the superposition of the textures of the subsystems. Direct DFT calculations on complete sliding trilayers confirm such superposition (Fig.~S26 in the SM). The three low-energy states exhibit multiple nonvolatile moir\'e polarization textures. Since the twist angles of $L_{12}$ and $L_{23}$ are exactly opposite, the atomic structures of $L_{12}$ and $L_{23}$ are symmetrically related, which imposes a strict transformation relationship between their polarizations ${\bf P}_{12}$ and ${\bf P}_{23}$. In the AA'-AA' state of twisted CrI$_3$ trilayer, the local sliding vectors ${\bf u}_{12}$ and ${\bf u}_{23}$ of $L_{12}$ and $L_{23}$ satisfy ${\bf u}_{23}$=$-{\bf u}_{12}$ [Fig.~\ref{Fig3}(a)]. The corresponding atomic structures are related through a $C_{2z}$ operation (i.e. $180^\circ$ rotation), consistent with the symmetry operation for sliding ferroelectricity in CrI$_3$ bilayers (Fig.~S6 in the SM). This local symmetry imposes a transformation relationship between the polarizations of $L_{12}$ and $L_{23}$: ${\bf P}_{12}$ and ${\bf P}_{23}$ carry opposite in-plane component ($P_{12,xy}=-P_{23,xy}$), but identical out-of-plane components ($P_{12,z}=P_{23,z}$). As shown in Fig.~\ref{Fig3}(b), the polarization distribution in AA'-AA' state is fully consistent with this quantitative constraint. This symmetry relationship is further validated by comparing the symmetry-derived ${\bf P}_{23}$ with direct DFT calculations over the sliding space (Fig.~S25 in the SM). It is worth noting that this $C_{2z}$ operation is applied locally, rather than to the entire polarization texture. For example, a local $180^\circ$ rotation transforms a center-convergent vortex into a center-divergent one and vice versa, whereas a global rotation preserves the vortex type. This local transformation rule directly governs the interference behavior of textures between the two subsystems in the twisted CrI$_3$ trilayer.

The twisted CrI$_3$ bilayers host vortex-antivortex polarization textures [Fig.~\ref{Fig3}(a) and Fig.~S12 in the SM]. The vortices with winding number $1$ are located at the AB'/BA' nodes and the antivortices with winding number $-2$ are located at the AA' nodes \cite{PRL-2025-CrI3-Twisted}. In the twisted trilayer, the relative moir\'e phase determines the superposition modes of these vortices and antivortices. As aforementioned, in the AA'-AA' state, the polarizations of the subsystem satisfy $P_{12,xy}=-P_{23,xy}$ and $P_{12,z}=P_{23,z}$, resulting in a net out-of-plane component only. As shown in Fig.~\ref{Fig3}(a), the net polarization texture (${\bf P}_{\rm total}$) of the AA'-AA' state forms alternating triangular domains of $P_z$, without global (anti)vortices anymore.

On the other hand, the AA'-BA' and AA'-AB' states of twisted CrI$_3$ trilayer involve nonzero relative moir\'e phases between $L_{12}$ and $L_{23}$, breaking the local $C_{2z}$ symmetry. Accordingly, their net polarization textures come from the constructive and destructive interference of (anti)vortices between $L_{12}$ and $L_{23}$. In the AA'-BA' state, the moir\'e phase shift aligns the center-convergent vortices of $L_{12}$ and $L_{23}$ [Fig.~\ref{Fig3}(c)]. The remaining center-convergent vortices and antivortices align with each other and only partially compensate the in-plane component. The antivortices are suppressed by such destructive superposition, forming the center-convergent-vortex lattice in the net polarization texture. In the AA'-AB' state, the center-divergent vortices are superposed, yielding a center-divergent vortex lattice [Fig.~\ref{Fig3}(d)]. These two distinct vortex textures, together with the non-vortex texture of the AA'-AA' state, establish a ternary system for moir\'e switching in the twisted CrI$_3$ trilayer. The continuous switching among these distinct polarization textures is further confirmed by mechanically driven sliding simulations (see Fig.~S17 in the SM and Video\_S1). These polarization textures can be experimentally resolved by lateral and vertical piezoresponse force microscopy, as demonstrated in twisted BN and WSe$_2$ \cite{NatCommun-2025-tBN-TopologicalPolarTexture, NanoLett-2024-WSe2-Texture}.

\begin{figure*}
	\centering
	\includegraphics[width=0.95\textwidth]{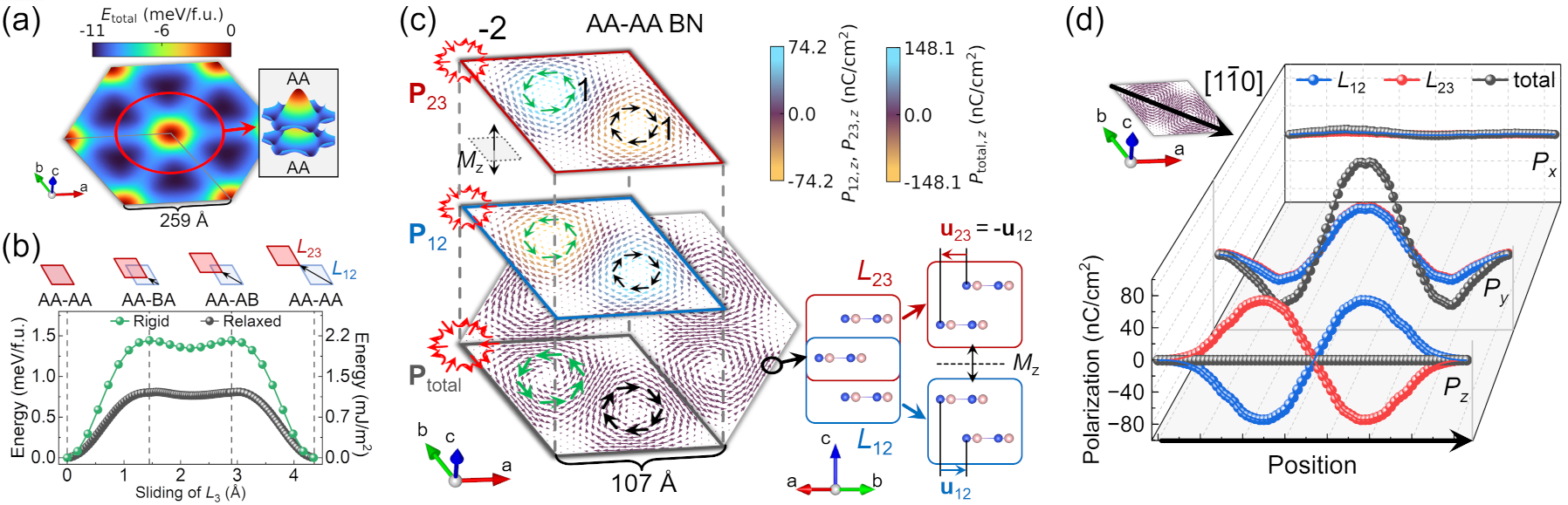}
	\caption{Results of twisted BN trilayer.
		(a) Local total energy ($E_{\rm total}$) landscape 
		(b) Switching barriers along one complete sliding period of $L_{3}$ obtained from rigid and relaxed sliding calculations. The schematics above illustrate the variation of the relative moir\'e phase between $L_{12}$ and  $L_{23}$.
		(c) Subsystem-resolved and net polarization textures of the low-energy AA-AA state. Inset: the $M_z$ symmetry relationship between $L_{12}$ and $L_{23}$. The arrow color represents the out-of-plane polarization $P_{z}$, while the arrow direction and length represent the direction and magnitude of the in-plane polarization.
		(d) Polarization components of ${\bf P}_{12}$, ${\bf P}_{23}$, and ${\bf P}_{\rm total}$ along the [1$\overline{1}$0] path indicated in the inset.
		For better visualization, the twist angles are $0.56^\circ$ in (a-b) and $1.35^\circ$ in (c-d).}
	\label{Fig4}
\end{figure*}

Now we turn to study the BN trilayer, which shows distinct behaviors. Unlike CrI$_3$, BN bilayer contains only one Mexican-hat-like profile at the AA node [Fig.~\ref{Fig1}(d)]. Consequently, the twisted BN trilayer only supports the AA-AA state with nested Mexican-hat-like energy landscape as the low-energy state [Fig.~\ref{Fig4}(a)]. As shown in Fig.~\ref{Fig4}(b), lateral sliding of $L_3$ drives the system through the unstable AA-BA and AA-AB states. For the $0.56^\circ$ twisted BN trilayer, such interlayer sliding can be mechanically driven by a lateral force of $0.013$ nN (Fig.~S15 in the SM), which is within an experimentally accessible range \cite{NatCommun-2025-tBN-MechanicalSliding}.

Although the twisted BN trilayer shows structural simplicity, its polarization texture remains nontrivial. As shown in Fig.~\ref{Fig4}(c), local $M_z$ operation connects $L_{12}$ and $L_{23}$ in twisted BN trilayer. Then the local sliding vectors satisfy ${\bf u}_{23}=-{\bf u}_{12}$. In twisted BN bilayers, antivortices with winding numbers $-2$ appear around the AA node. More interestingly, polarization merons and antimerons with winding number $1$ appear around the AB and BA nodes, respectively. As quantitatively illustrated in Fig.~\ref{Fig4}(d), the local $M_z$ symmetry preserves the in-plane polarization ($P_{12,xy}=P_{23,xy}$), while reversing the out-of-plane component ($P_{12,z}=-P_{23,z}$) in the AA-AA state of twisted BN trilayer, in contrast to the AA'-AA' state of twisted CrI$_3$ trilayer. This symmetry relationship is also validated by DFT calculations (Figs.~S24 and S25 in the SM). Consequently, the net polarization no longer takes the form of the meron-antimeron texture as in the twisted BN bilayer, but degenerates into a vortex-antivortex texture.

Finally, it should be noted that the AA-BA and AA-AB states of twisted BN trilayer can host trimeron-like polarization textures (Fig.~S23 in the SM). However, they are not energetically stable by themselves. Due to the similarity in stacking-energy landscapes between BN and WSe$_2$ \cite{AdvMater-2025-WSe2-Terraced}, our results offer insights into the experimental observations of strain-free twisted WSe$_2$ trilayers, whose moir\'e polarization texture only exhibits volatile responses \cite{NatNanotechnol-2024-WSe2-3L}.

In summary, we have established a general theoretical scenario for moir\'e polarization textures in twisted trilayers. Beyond the bilayer limitation, the relative moir\'e phase between these two subsystems serves as an additional structural degree of freedom to control the global stacking landscape and the associated polarization textures. The existence, stability, and morphology of polarization textures in trilayers can be inferred from the stacking-energy landscape and symmetry-governed polarization texture of corresponding bilayers. The resulting trilayer polarization textures remain robust against structural relaxation, uniaxial strain, and thermal perturbations (Figs.~S19-S21 in the SM, Video\_S3 and Video\_S4). Both CrI$_3$ and BN have been studied as model systems, which exhibit highly contrastive behaviors regarding the moir\'e polarization textures and their switchability. Our results provide a general route toward engineering multiple polarization textures in twisted moir\'e superlattices.

\begin{acknowledgments}
	This work was supported by National Key Research and Development Program of China (Grant No. 2025YFA1411100), National Natural Science Foundation of China (Grants No. 12325401 and No. 12274069) and the Big Data Computing Center of Southeast University.
\end{acknowledgments}
\bibliography{Refs}

\end{document}